\documentstyle[aps,prd,preprint]{revtex}

\begin{document}

\title{Open inflation and the singular boundary}
\author{Jaume Garriga\footnote{Electronic address: 
garriga@ifae.es}}

\address{IFAE, Departament de Fisica,\\
         Universitat Autonoma de Barcelona,
         08193 Bellaterra, Spain.}
\date{\today}
\maketitle

\begin{abstract}
The singularity in Hawking and Turok's model of open inflation 
has some appealing properties. We suggest that this singularity 
should be regularized with matter. The singular instanton
can then be obtained as the limit of a family of ``no-boundary''
solutions where both the geometry and the scalar field are regular. 
Using this procedure, the contribution of the singularity to the 
Euclidean action is just $1/3$ of the Gibbons-Hawking boundary term. 
Unrelated to this question, we also point out that
gravitational backreaction improves the behaviour of 
scalar perturbations near the singularity. As a result,
the problem of quantizing scalar perturbations and gravity waves 
seems to be very well posed.
\end{abstract}
\vskip 2 truecm

Recently, Hawking and Turok \cite{HT,HT2} have suggested
that an open universe can be created from nothing. 
This is an attractive possibility because it would allow 
to construct open models of inflation with very simple 
inflationary potentials (see also \cite{everybody,alex,unruh}). 

The new ingredient that makes their construction possible
is that they allow their instanton solution to be singular. 
There is some justification for this, since
the Euclidean action is integrable near the singularity.
Moreover, if we think of the singularity as the boundary of
spacetime, the Gibbons-Hawking boundary term \cite{GH}
is non-vanishing and finite. This is rather 
coincidental, since it requires
the extrinsic curvature of the boundary to increase just at
the same rate as the inverse of its volume as the singularity is 
approached. 

In this paper, we suggest that the singularity should be regularized
with matter, so that the instanton can be obtained as the limit of
a family of nonsingular geometries where the scalar field is
also well behaved. The simplest way to do this is to introduce a membrane
coupled to the scalar field. The Euclidean action
is given by
\begin{equation}
S_E= \int d^4 x \sqrt{g}
\left[{1\over 2}(\partial \phi)^2 +V(\phi) - {{\cal R}\over 16\pi G} \right] +
\int d^3 \xi \sqrt{h} \mu(\phi).
\label{action}
\end{equation}
Where 
\begin{equation}
\mu(\phi) = \mu_0 - \alpha e^{\kappa \phi},
\label{coupling}
\end{equation}
and $h$ is the determinant of the metric on the worldsheet of the
membrane. The parameter $\mu_0>0$ is a positive tension which stabilizes 
the vacuum at $\phi=0$, and $\alpha$ is a small coupling. These parameters 
will not play a role once the ``singular'' limit is taken, but for the 
time being there is no harm in thinking of them as physical. The
parameter $\kappa$ will be specified below.
We have not written a boundary term, since our geometries will not 
have a boundary.

Following \cite{HT} we take an O(4)-symmetric ansatz for the 
metric and the scalar field:
\begin{equation}
        ds^2 = d\sigma^2 + b^2(\sigma) (d\psi^2 
                + \sin^2\psi d\Omega_2^2).
\label{eq:metric}
\end{equation}
In the absence of a membrane,,
the field equations for $b(\sigma)$ and $\phi(\sigma)$ are
\begin{equation}
        \phi'' + 3{b' \over b}\phi' = V_{,\phi}\,,
\label{eq:phieq}
\end{equation}
\begin{equation}
        \left({b'\over b}\right)^2= 
{8\pi G \over 3} \left[{1\over 2}\phi'^2- V\right] + {1\over b^2},
\label{eq:beq}
\end{equation}
where primes stand for derivatives with respect to $\sigma$.

The instanton is regular at $\sigma=0$, where $b\approx \sigma$ and
$\phi'=0$. As $\sigma$ is increased, $b$ grows to a maximum value and
then decreases again, reaching a secon zero at some $\sigma=\sigma_f$. 
However, this second zero is singular. The scale factor there behaves
as \cite{HT,alex}
$$
b^3 \approx C (12\pi G)^{1/2}(\sigma_f-\sigma)
$$
and the scalar field as
$$
\phi \approx -(12 \pi G)^{-1/2} \ln(\sigma_f-\sigma)+ const.
$$
In spite of the singular behaviour of the scalar field and the
geometry, the Euclidean action is integrable. 

Here, we shall take the approach of modifying the solution
so that it will be everywhere regular.
The idea is to surround the singularity with a spherical 
membrane which will act as a source for the scalar field.
The interior of the membrane is replaced with a ball of
(nearly) flat space. At the center of the ball, $\sigma=\sigma_c$,
we take $\phi'=0$, $b'=-1$, and $\phi(\sigma_c)$ is chosen so that 
it matches the value of $\phi$ at the membrane. The membrane 
will also provide the energy momentum source necessary to match
both geometries.

Substituting the O(3) symmetric ansatz into the Euclidean action
and varying with respect to $\phi$, one easily finds 
the matching conditions for the scalar field at the membrane.
The discontinuity in the first derivative is given by
\begin{equation}
\label{match}
[\phi'(\sigma_m)] = -\alpha \kappa e^{\kappa \phi(\sigma_m)},
\end{equation}
where the square brackets indicate the difference between the 
values inside and outside, and $\sigma_m$ is the location
of the membrane. given that $\phi'\approx 0$ inside
the membrane and using the asymptotic form of $\phi'$ near the
external face we have
\begin{equation}
{C\over (12 \pi G)^{1/2}} \approx 
\alpha b^3(\sigma_m) e^{\kappa \phi(\sigma_m)}.
\label{1}
\end{equation}
The left hand side of this equation is constant. In
order to obtain a nontrivial limit as $\sigma_m\to \sigma_f$ while keeping
$\alpha$ finite we take
\begin{equation}
\kappa \equiv (12\pi G)^{1/2}.
\label{kappa}
\end{equation}

Let us now consider the backreaction of this membrane on the
geometry. Einstein's equations imply the matching condition
\cite{israel}
\begin{equation}
\left[{b'\over b}\right]= -4\pi G \mu(\phi) = 
-4 \pi G (\mu_0 - \alpha e^{\kappa \phi(\sigma_m)}).
\label{matching}
\end{equation}
Inside the membrane, the geometry is basically flat, and
we have $(b'/b)\approx b^{-1}$. Outside the membrane, we have
\begin{equation}
{b' \over b} \approx  {-\kappa C \over 3 b^3}.
\label{2}
\end{equation}
Using (\ref{1}) we find that the leading $O(b^{-3})$
terms in (\ref{matching}) cancel out. The subleading terms 
are unimportant; they will not contribute once the size of 
the membrane is shrunk to zero.

Inserting the trace of Einstein's equations in (\ref{action}), we find
\cite{bt}
\begin{equation}
S_E= - \int d^4 x \sqrt{g} V(\phi) - 
{1\over 2}\int d^3 \xi \sqrt{h} \mu(\phi).
\label{action2}
\end{equation}
The limit of the second term as the size of the membrane is shrunk to
zero can be interpreted as the contribution of the 
singularity to the action of the instanton. It is given by
\begin{equation}
S_{sing} = {\pi^2 C \over \kappa}.
\label{boundary}
\end{equation}
Taking into
account that the trace of the extrinsic curvature of the membrane
is $K= 3 (b'/b)$, and using (\ref{2}), we find that this contribution 
is actually one third of the Gibbons-Hawking term \cite{GH,alex,HT2}
evaluated on the external face of the membrane
\begin{equation}
S_{sing}={1\over 3} S_{GH}={- 1 \over 24 \pi G} \int d^3 \xi \sqrt{h} K_{ext}.
\label{sing}
\end{equation}
This conclusion is rather general. The junction condition \cite{israel}
$[K]=-12 \pi G \mu$ relates the value of
$\mu(\phi)$ in Eq. (\ref{action2}) to the jump in the trace of the 
extrinsic curvature. However, the jump in $K$ is dominated
by the extrinsic curvature on the external face,
from which (\ref{sing}) follows.

Note that the result in (\ref{boundary}) does not depend on the 
parameters $\mu_0$ or $\alpha$ characterizing the membrane. The reason is
that $\alpha$ has been elliminated in favour of $C$ through equation (\ref{1}),
whereas $\mu_0$ does not contribute in the limit $b(\sigma_m)\to 0$. In fact,
there is no strong reason for using a coupling of the form
(\ref{coupling}). It has been chosen so that the regulator
$\alpha$ remains finite as the singularity is approached.\footnote{
We could replace $\mu_0$ by $\mu_0+ \beta e^{\kappa/3}\phi_0$, and then 
$\mu_0$ and $\beta$
would also remain finite in the limit $b(\sigma_m)\to 0$.}
If we think of our membrane as a physical object,
then for each $C$ and for each value of the cut-off $\sigma_m$,
the solution only exists for specific values of $\mu_0$ and 
$\alpha$ determined by the matching conditions. One can extend this 
interpretation by taking the coupling $\alpha$ very small and allowing 
for a superposition of 
any number of membranes with positive and negative charges. In this case, 
the parameters $\mu_0$ and $\alpha$ can be thought of
as continuous variables, which can be adjusted to satisfy (\ref{1})
for any value of $C$ and $\sigma_m$.

The instability of flat space pointed out by
Vilenkin \cite{alex} can be seen as the spontaneous creation of a membrane 
which is a source for the scalar field. Because $\phi$ is large near the 
membrane, its
effective energy per unit area $\mu(\phi)$ is negative. This negative
energy compensates for the positive energy in the scalar field
configuration, so that the total energy is zero and tunneling is allowed.
In Ref. \cite{alex}, a massless scalar field was considered,
and there was no minimum gap to be surmounted in order 
for tunneling to occur (the constant $C$ could be chosen arbitrarily small).
This may also be true for a generic potential, and in this case
it seems that the same regularization that makes the Hawking-Turok
instanton acceptable also makes flat space unstable. 
There may be models, however, where there is a minimum height of the
tunneling barrier. These models would make flat space metastable at least. 

The solution of Hawking and Turok is also special with regard
to the unrelated question of cosmological perturbations.
In the approximation 
when the gravitational backreaction of
the scalar field perturbations is neglected,
Hawking and Turok \cite{HT} have argued that
the quantization of fluctuations is marginally well defined 
in spite of the singularity. Indeed, after the rescaling $\phi=\chi/b$, 
and introducing the conformal coordinate 
$X=\int_{\sigma}^{\sigma_f} d\sigma/b(\sigma)$
the field modes obey a Schrodinger equation with a potential that
behaves as $-(2 X)^{-2}$ near the singularity. This is again
very coincidental, since with a stronger singularity 
the quantum mechanical problem would certainly be ill posed \cite{HT,gapa}.

Therefore it is important to check what happens when gravitational 
backreaction is included. 
The quantization of cosmological perturbations in 
$O(3,1)$ symmetric geometries (the analytic continuation of our instanton
has this symmetry) was recently studied in Ref. \cite{GMTS}.
The analysis is not straightforward because the $\sigma=const.$ 
surfaces of homogeneity and isotropy cannot be used as cauchy surfaces
on which commutation relations can be imposed. Instead, one has
to resort to inhomogeneous $\psi=const$ surfaces, where the 
disentanglement of scalar and tensor modes is complicated. 
However, the final result is
rather simple. After a suitable rescaling \cite{GMTS}, 
the gauge invariant
scalar potential obeys a Schrodinger equation with effective potential 
given by
\begin{equation}
{4 \pi G} \phi'^2 + \phi'\left({1\over \phi'}\right)''.
\label{effective}
\end{equation}
Here, a prime indicates derivative with respect to the conformal coordinate
$X$ introduced above. It is straightforward to show that the first term
dominates near the singularity, behaving as $k/X^{2}$, with $k=3/4$.
Hence, the effective potential goes to plus infinity rather than minus 
infinity near the singularity. 
Interestingly, the coefficient $k=3/4$ is again
a critical one \cite{gapa}. As mentioned above, for $k < -1/4$ the
problem is not well posed. For $-1/4 < k < 3/4$ the problem is 
marginally well posed, since both solutions of the Schrodinger
equation are square integrable near the singularity, but only one
has a square integrable kinetic energy. Finally, for $k \geq 3/4$,
the basis of functions is uniquely determined by the requirement of
square integrability \cite{gapa}, which selects one solutions
for each value of the energy. Thus, the problem of quantizing the
perturbations seems much better posed thanks to backreaction.
In particular, this seems to preclude the possibility of matter
``streaming out'' from the singularity into the universe \cite{unruh}.
The same comment applies to gravity waves, for which the 
corresponding effective potential reduces to the first term in
(\ref{effective}) \cite{GMTS}.

It is a pleasure to thank Alex Vilenkin Takahiro Tanaka and 
Xavier Montes for very useful conversations.

\end{document}